\documentstyle[preprint,aps]{revtex}

\title{The asymptotic dynamics of three-dimensional Einstein
 gravity with a
negative cosmological constant }

\author{Oliver Coussaert$^{(a)}$, Marc Henneaux$^{(a,b)}$ and Peter
 van Driel$^{(a)}$}

\address{$^{(a)}$ Universit{\'e} Libre de Bruxelles\\
         CP 231, Bvd du Triomphe, B-1050 Brussels, Belgium\\
$^{(b)}$ Centro de Estudios Cient\'{\i}ficos de Santiago,\\
Casilla 16443, Santiago 9 Chile\\
         }

\begin{document}

\maketitle
{\flushright ULB-TH/95/08}

\begin{abstract}Liouville theory is shown to describe the asymptotic
 dynamics of
three-dimensional  Einstein gravity with a negative cosmological
constant. This is because (i) Chern-Simons theory with a gauge group
$SL(2,R)  \times SL(2,R)$ on a space-time with a cylindrical boundary
is equivalent to the non-chiral $SL(2,R)$ WZW model; and (ii) the
anti-de Sitter boundary conditions implement the constraints that
reduce the WZW model to the Liouville theory.
\end{abstract}

\section{Introduction}

Three-dimensional gravity theories have attracted considerable
attention in the past fifteen years in the hope of getting a better
understanding of the intricacies of their four-dimensional parents
 (see \cite{1,2}
and \cite{3} for a recent review with an extensive list  of
references). In particular, the asymptotic structure of 3d gravity and
3d supergravity has been investigated in \cite{4,5,6}, with the
following conclusions.
In the case of a vanishing cosmological constant, the asymptotic
behaviour of the gravitational field is quite constrained and does not
allow one to define naturally translations and supersymmetries at
spatial infinity. [This property has been argued recently to explain
how one could break supersymmetry without generating a cosmological
constant \cite{7}]
By contrast, the asymptotic structure of 3d gravity with a negative
cosmological constant $\Lambda<0$, is much richer\footnote{The
assumption made in the footnote 11 of \cite{4}, that the contraction from
$\Lambda<0$ to $\Lambda=0$ can be done smoothly, is thus incorrect.}.

Indeed, in that case the asymptotic symmetry group turns out to be the
group of conformal transformations in two dimensions, generated by
the infinite dimensional Virasoro algebra. The appearance of this
conformal symmetry can be understood either in terms of the Penrose
description of infinity by means of a conformal compactification,
where infinity appears as a timelike cylinder ($\Lambda<0$) and the
asymptotic symmetry group is the group of its conformal
symmetries \cite{8} ; or in terms of the Hamiltonian formulation where
the canonical generators of the transformations preserving the boundary
conditions are shown to close according to the conformal algebra \cite{6}.

The presence of the infinite-dimensional conformal group as asymptotic
symmetry group strongly suggests that the asymptotic dynamics of the
gravitational field in 3 dimensions, with a negative cosmological
constant, is described by a two-dimensional conformal field theory.
The purpose of this letter is to show that the conformal field theory in
question is Liouville theory \footnote{That there is a connection
  between three-dimensional gravity and Liouville theory is of course
  not new and has been discussed from a different perspective in
\cite{9}
 and \cite{10}.}.

Our starting point is the crucial observation made in \cite{11,12}  that
three-dimensional Einstein gravity with $\Lambda<0$ can be
reformulated as a Chern-Simons gauge theory with gauge group $SL(2,R)
\times SL(2,R)$ and action
\begin{equation}
\label{1}
S_E[A,{\tilde A}]=S_{CS}[A]-S_{CS}[{\tilde A}].
\end{equation}
Here $A$ (respectively $\tilde A$) is the gauge field associated with
the first
(respectively the second) $SL(2,R)$ factor and $S_{CS}$ is the
Chern-Simons action which  in  polar coordinates $t,r, \varphi$ takes
 the form,
\begin{equation}
\label{2}
S_{CS}=\int dt dr d\varphi \ \ tr({\dot A}_r
 A_\varphi-{\dot A}_\varphi A_r - A_0 F_{r \varphi})
\end{equation}
The connections $A$ and $\tilde A$ are related to the triad $e$ and spin
 connection $\omega$ through $A=e+\omega$, $\tilde A = e- \omega$.

The surface at spatial infinity ($r=\infty$) of anti-de-Sitter space is a
timelike cylinder with coordinates $t,\varphi$. We denote it by
$\Sigma_2$. The boundary conditions on the metric given in~\cite{6} read,
 when translated in terms of the connection $A$ and $\tilde A$,
\begin{equation}
\label{3}
A \sim \left[
\begin {array}{cc} \frac{dr}{2r}& O(1/r)\\
\noalign{\medskip}{r dx^+} & -\frac{dr}{2r}
\end{array}
\right],  \qquad
{\tilde A} \sim \left [
\begin {array}{cc} -{\frac {dr}{2r }}&r dx^-\\
\noalign{\medskip}O(1/r)&{\frac {dr}{2r }}
\end{array}
\right]
\end{equation}
(to leading order). Here, $x^{\pm}=t \pm \varphi$. The boundary
conditions express that the metric approaches asymptotically the
anti-de Sitter space and are in particular such that the triad $e$ is
non degenerate.

Two things should be emphasised about~(\ref{3}). (i) First, the lightlike
components $A_-$  of $A$ and $\tilde A_+$ of $\tilde A$ are set equal
to zero asymptotically. (ii) Second, $A^{(-)}_+$ and ${\tilde A}^{(+)}_-$
are not functions of the variables $t$ and $ \varphi$ to leading order in $r$.
At the same time, $A^{(3)}_+$ and ${\tilde A}^{(3)}_-$ are set to vanish.
 Here, the indices in parentheses are Lie algebra indices.
We shall examine in turn the respective implications of (i) and (ii).
We start by showing that (i) reduces the Chern-Simons theory to the
$SL(2,R)$ non-chiral Wess-Zumino-Witten model. To that end, we closely
follow the work of~\cite{13}, adapted to the boundary conditions at hand.
In section III we show than that the implication of (ii) is to reduce
 this WZW model to the Liouville model.

\section{From the Einstein action to the non-chiral $SL (2,R)$the case
 Wess-Zumino-Witten model}

The action  (\ref{1}) is not an extremum on-shell when $A_-$ and
$\tilde A_+$ are required to vanish on the boundary. Rather, $\delta
S$ is then equal to the surface term  $ \delta[ \int_{\Sigma_2} dt
d\varphi  \  \ tr(A^2_\varphi + {\tilde A}^2_\varphi)]$ on the surface
$\Sigma_2$ at infinity. [ We shall examine the terms that arise at
$t_1$ and $t_2$ when discussing (ii)]. In order to have $\delta S=0$,
one must therefore add to the action  the surface term   $- [
\int_{\Sigma_2} dt d\varphi \ \ tr(A^2_\varphi + {\tilde
  A}^2_\varphi)]$, leading to the improved action
\begin{eqnarray}
\label{4}
S[A,{\tilde A}] = S_{CS}[A] - \int_{\Sigma_2} dt d\varphi
tr(A^2_\varphi) - S_{CS}[{\tilde A}] - \int_{\Sigma_2} dt d\varphi
 tr({\tilde A}^2_\varphi)
\end{eqnarray}

The temporal component $A_0$ and $\tilde A_0$ of the vector potential
appears as Lagrange multiplier implementing the constraints $
F_{r\varphi}={\tilde F}_{r\varphi}=0$. One can solve these constraints
as
\begin{equation}
\label{5} A_i=G^{-1}_1 \partial_i G_1, \ \  {\tilde
    A}_i=G^{-1}_2 \partial_i G_2
\end{equation}
where $G_1$ and $G_2$ are asymptotically given by
\begin{equation}
\label{6}
G_1 \sim g_1(t, \varphi) \left [\begin {array}{cc}
 \sqrt {{r}}&0\\\noalign{\medskip}0&{
\frac {1}{\sqrt {{r}}}}\end {array}\right ], \qquad
G_2 \sim g_2(t, \varphi) \left [\begin {array}{cc}
 \frac {1}{\sqrt{r}}&0\\\noalign{\medskip}0&\sqrt {{r}}\end {array}\right ]
\end{equation}
and where  $g_1(t,\varphi)$ and $g_2(t,\varphi)$ are arbitrary
elements of $SL(2,R)$. With (\ref{6}), the asymptotic behaviour of the radial
components of $A$ and $\tilde A$ coincide with the one of (\ref{3}) ,
while the tangential components behave as
\begin{eqnarray}
\label{7}
A_\alpha \sim  \left [\begin {array}{cc} a^{(3)}_\alpha &
\frac{a^{(+)}_\alpha }{r}  \\\noalign{\medskip}
 {a^{(-)}_\alpha}{r} & - a^{(3)}_\alpha \end {array}\right ], \qquad
{\tilde A}_\alpha \sim  \left [\begin {array}{cc} {\tilde a}^{(3)}_\alpha &
   {\tilde a}^{(+)}_\alpha r  \\\noalign{\medskip}
 \frac{{\tilde a}^{(-)}_\alpha}{ r} & - {\tilde a}^{(3)}_\alpha \end
{array}\right ]
\end{eqnarray}
where   $a_\alpha=g^{-1}_1 \partial_\alpha g_1$ and ${\tilde
  a}_\alpha=g^{-1}_2\partial_\alpha g_2  $. The group elements
$g_1(t,\varphi)$ and $g_2(t,\varphi)$ will be restricted below so that
$A_\alpha$ and ${\tilde A}_\alpha$ fulfill the remaining boundary
conditions (ii).

Strictly speaking~(\ref{5}) is valid only if the spatial sections
have no hole. In general, one should allow for holonomies, which
appear as additional "zero mode terms" in (\ref{5}) . Such additional
terms are necessary to describe  black holes in 3 dimensions
which can be obtained from anti-de Sitter space by making appropriate
identificationsxi \cite{14,15}. Furthermore, there are then also additional
inner
boundaries with their own surface dynamics. The dynamics on a black
hole horizon,
 has been treated in \cite{16}.
Since we are interested here only in the asymptotic dynamics of
 the gravitational field,
 we shall however drop the holonomies and ignore the inner surfaces.
 A full treatment
 will be given in  \cite{17}.

Now, if one inserts~(\ref{6}) in the action~(\ref{5}), one gets
 \begin{equation} \label{8}  S[A,{\tilde A}] = S^R_{WZW}[g_1] - S^L_{WZW}[g_2]
 \end{equation} where   $S^R_{WZW}[g_1]$ and $ S^L_{WZW}[g_2]$ are
 the two-dimensional
 chiral Wess-Zumino-Witten (WZW) actions  \cite{13,18,20,20b,21}.
These first order action

 generalise the abelian actions of \cite{19}  and respectively describe a
 right-moving group element $g_1(x^+)$ and a left moving group element
$g_2(x_-)$,
 \begin{eqnarray}
\label{9}
S^R_{WZW}[g_1]=\int_{\Sigma_2} dt d\varphi \, tr({\dot g_1}
{g'_1}-{(g'_1)^2})+\Gamma[g_1]  \\ \label{10} S^L_{WZW}[g_2]
=\int_{\Sigma_2} dt d\varphi \, tr({\dot g_2}
{g'_2}+{(g'_2)^2})+\Gamma[g_2]
                 \end{eqnarray}
 where ${\dot g}=g^{-1} \frac{\partial }{\partial t} g$,  ${ g'}=g^{-1}
\frac{\partial }{\partial \varphi} g$ and $\Gamma[g]$ is the usual
three-dimensional part of the WZW-action.
As shown in  \cite{18,20,21}, the actions (\ref{9}) and (\ref{10})
each lead to a single chiral Kac-Moody symmetry (of opposite
chirality). One expects the sum (\ref{8}) of the left and right
chiral actions (\ref{9} -\ref{10}) to be equivalent to the standard,
non chiral, WZW action \cite{22} with dynamical
variable $g=g^{-1}_1 g_2$ since in that model the right
moving and left moving sectors are indeed decoupled \cite{22,22b}.
This expectation  turns out to be true.

One to establish the equivalence is to rewrite the standard WZW
action in Hamiltonian form, since  (\ref{9}) and (\ref{10}) are linear
and of first order in the time derivatives. We denote by $\Pi_g$ the
momentum conjugate to $g$ and by $u$ the function of $g$ and $\Pi_g$
which is equal to $\dot g$ when the equations of motion hold. One may
take $g$ and $u$ as independent variables. The change of variables
\begin{equation}
\label{11}
g=g^{-1}_1 g_2, \ \ \ u \equiv {\dot g}|_{on shell} =-g^{-1}_2
\frac{\partial}{\partial \varphi} g_1 g^{-1}_1 g_2 - g^{-1}_2
\frac{\partial}{\partial \varphi}g_2
\end{equation}
brings (\ref{8}) to the standard WZW action in first order form
or, after elimination of the auxiliary field $u$, to the standard, non
chiral $SL(2,R)$ WZW action in second order form,
\begin{eqnarray}
\label{12}
S[A, {\tilde A}] = S^{WZW}[g],\ \ \
S^{WZW}[g]=\int_{\sigma_2} dt d\varphi  (tr( g_+ g_-) - \Gamma[g])
\end{eqnarray}
where $g_\pm \equiv  g^{-1} \frac{\partial}{\partial x^\pm} g$.
We omit the details here, leaving them for the complete treatment
\cite{17}  where, in particular the zero modes and holonomies will be
 included also. We simply note that the transformation (\ref{11})
is the direct generalisation of the transformation analysed in
\cite{18}, \cite{23} in the $U(1)$ case, which  establishes the
equivalence of the sum of left moving chiral boson and a right moving
 chiral boson
to a massless Klein-Gordon field.

We have thus shown so far that the asymptotic dynamics of the
gravitational field in three dimensions with $\Lambda<0$ is described
by the (non-chiral) $SL(2,R)$ WZW action. We have not yet
incorporated, however, all the boundary conditions on the connection.
This missing step is taken now.

\section{From the WZW model to Liouville theory}

The conditions that we have not taken into account at this stage are
the conditions (ii) which read, in terms of the group element $g$
\begin{equation}\label{constraints}
J^{(+)}_-\equiv (g^{-1} \partial_- g )^{(+)} = 1,\qquad
{\tilde J}^{(-)}_+ \equiv ( \partial_+ g g^{-1})^{(-)} =  1
\end{equation}
and $J^{(3)}_+=0, \ J^{(3)}_-=0$. Since the $J$'s are just the
Kac-Moody currents of the WZW model, we recognise (\ref{constraints})
 as the conditions
implementing the familiar Hamiltonian reduction of the WZW model to
the Liouville theory. The conditions $J^{(3)}_+= J^{(3)}_-=0$ appear
as ``gauge condition''. This reduction has been discussed at length in
the literature  so that we do not need to recall  the details here.

Let us simply point out the perhaps less familiar fact that the
reduction can be carried out directly at the level of the action. As
shown in \cite{24} , the WZW action reads, if one parametrises $g$
 according to
the Gauss decomposition
\begin{eqnarray}
g=\left [\begin {array}{cc} 1&X\\\noalign{\medskip}0&1\end {array}
\right ] \left [\begin {array}{cc}
    \exp(\frac{1}{2}\phi)&0\\\noalign{\medskip}0&\
exp(-\frac{1}{2}\phi)\end {array}
\right ]  \left [\begin {array}{cc} 1&0\\\noalign{\medskip}Y&1\end {array}
\right ] , \\
S^{WZW}[g]= \int dt d\varphi [ \frac{1}{2} \partial_+ \phi \partial_-
\phi + 2 (\partial_- X) (\partial_+ Y) \exp(-\phi)] \label{err}
\end{eqnarray}

Now the action (\ref{12}) is defined on the cylinder $[t_1,t_2] \times
 S_1$ of finite
height $t_2-t_1$ and is stationnary on the classical history provided
one fixes $\phi, X$ and $Y$ at the time boundary  $t_1$ and $t_2$.
 However, since we want
 to implement the constraints (\ref{constraints}),
it is not $\phi,X$ and $Y$ that we want to fix at the boundaries, but
rather, $\phi$ and the momentum $\partial_- Y$ and $\partial_+ X$
conjugate to $X$ and $Y$, since $J^{(+)}_+=\partial_- X \exp(-\phi)$
and ${\tilde J}^{(-)}_-=\partial_+ Y \exp(-\phi)$.
Hence the constraints~(\ref{constraints}) cannot be simply plugged in
inside~(\ref{err}). The action appropriate to the new set of
boundary conditions differs from~(\ref{err}) by a boundary term at
 $t_1$ and $t_2$,
\begin{equation}
\label{16}
S^{WZW}_{impr}[g] = S^{WZW}[g] - 2\oint
d\varphi  (X \partial_+ Y + Y \partial_- X) \exp(-\phi)|^{t_2}_{t_1}
\end{equation}
With the ``improved'' term,  it is legitimate to insert the
 constraints~(\ref{constraints}) in
the action~(\ref{16}). If one does so, one ends up with the
 Liouville action for
$\phi$
\begin{equation}
\label{17}
S[A,{\tilde A}] = S_{Liouville}[\phi]=\int dt d\varphi
 (\frac{1}{2} \partial_+ \phi \partial_- \phi + 2 \exp(\phi))
\end{equation}

Hence, we have established that the asymptotic dynamics of the
gravitational field in three dimensions, with $\Lambda<0$, is
indeed described by the
Liouville theory. As it is well known, this theory is conformally
invariant and possesses two sets of Virasoro generators $L_n$ and
$\tilde L_n$.  These can be viewed as generating the residual
Kac-Moody symmetries preserving the constraints and are the asymptotic
generators found by a totally different approach in \cite{6}. [See
also \cite{11p} and \cite{11pp} for a related analysis of the surface
terms in the Chern-Simons theories.]

Remark: one can actually substitute the constraints $ J^{(+)}_+=\mu$ and
${\tilde J}^{(-)}_-=\nu$ in the action (\ref{16}), provided one
observes that the constants $\mu$ and $\nu$ are functionals of the
fields and varies them accordingly in the action principle. A very similar
situation occurs when one treats the cosmological constant as a
dynamical variable. The subtleties of the variational principle are
explained in that case in \cite{25}.

\section{Conclusions}

In this letter, we have completed the analysis of the asymptotic
dynamics of 3d Einstein gravity with a negative cosmological constant.
We have shown that the Virasoro symmetry generators found in \cite{6} arise
because the asymptotic dynamics is described by a conformally
invariant theory, namely the Liouville theory.

The asymptotic reduction of the Einstein action -equivalent to
$SL(2,R) \times SL(2,R)$ Chern-Simons action- to the Liouville action
follows a two-step procedure. First, one imposes conditions of
opposite chiralities on  each $SL(2,R)$ factor, namely $A_-=0$ and
${\tilde A}^+ =0$. This leads to the sum of two chiral $SL(2,R)$ WZW
actions of opposite chiralities or, what is the same, to the non
chiral $SL(2,R)$ WZW action. Next one imposes the constraints on
the Kac-Moody currents that reduce the $SL(2,R)$ WZW action to the
Liouville theory. The two steps are precisely incorporated in the
boundary conditions on the triads and connection  expressing
asymptotic
 approach
to the anti-de Sitter space, and thus, they have a direct geometrical
interpretation. Our analysis also exemplifies very clearly how
 the asymptotic dynamics is sensitive to the boundary conditions.

A further account of this work, with extension to supersymmetry
(important for proving positivity of the energy theorems) will
 be reported elsewhere.
\vskip 1 cm
\noindent {\bf Acknowledgements}.
\vskip 1 cm
\noindent One of us (M.H.) is grateful to Lars Brink and Steve Carlip for
discussions. This work has been partly supported by research funds
from the FNRS and the Commission of the European Communities. O.C. is
Aspirant FNRS.

\end{document}